\documentclass[aps,preprint,prd,epsfig,nofootinbib]{revtex4}
\pdfoutput=1
\usepackage{amssymb}
\usepackage{amsmath}
\usepackage{color}
\usepackage{fancyhdr}
\usepackage{slashed}
\usepackage[utf8]{inputenc}
\usepackage{graphicx}
\usepackage{epsfig}
\usepackage{amssymb}
\usepackage{bm}
\usepackage{color}
\usepackage[dvipsnames]{xcolor}
\newcommand{\be}{\begin{equation}}
\newcommand{\ee}{\end{equation}}
\newcommand{\bea}{\begin{eqnarray}}
\newcommand{\eea}{\end{eqnarray}}
\newcommand{\bes}{\begin{subequations}}
\newcommand{\ees}{\end{subequations}}

\newcommand{\bc}{\begin{center}}
\newcommand{\ec}{\end{center}}
\usepackage{amsmath, amssymb} 
\usepackage{slashed}
\usepackage{graphicx}
\begin{document}

\title{Evading the Landau pole in the minimal 3-3-1 model with leptoquarks}

\author{A. Doff$^{a}$, C. A. de S. Pires$^{b}$}
\affiliation{{$^a$ Universidade Tecnol\'ogica Federal do Paran\'a - UTFPR - DAFIS,  84017-220, Ponta Grossa, PR, Brazil}, \\
{$^b$  Departamento de Física, Universidade Federal da Paraíba, Caixa Postal 5008, 58051-970, Jo\~ao Pessoa, PB, Brazil}, \\
}

\date{\today}

\begin{abstract}
In its original version, the minimal 3-3-1 model possess a Landau-pole around 2-6 TeV scale. Current LHC bound on $Z^{\prime}$ implies that  the $SU(3)_L\times U(1)_X$ symmetry must break spontaneously around 4 TeV which means that the model may lose its perturbative character even before symmetry breaking. This is a disaster for the model. Few attention has been devoted to this problem.  Here we investigate the efficiency of scalar leptoquarks
in evading or shifting the Landau pole to a harmless energy scale.
\end{abstract}

\maketitle
\section{Introduction}
The gauge models for the electroweak and strong forces  based in the  gauge group $SU(3)_C\times SU(3)_L\times U(1)_X$ (3-3-1 )\cite{Singer:1980sw,Pisano:1992bxx,Frampton:1992wt,Foot:1994ym,Montero:1992jk} accommodate two versions determined by the parameter $\beta$ that appears in the linear combination of the diagonal generators that define the electric charge operator,
\begin{equation}
    \frac{Q}{e}=\frac{1}{2}(\lambda_3+\beta \lambda_8)+N.
\end{equation}
The values $\beta$ can take are $-\frac{1}{\sqrt{3}}$ (case A) and $-\sqrt{3}$ (case B). Each case leads to different models concerning theoretical and phenomenological aspects\footnote{It is interesting to stress that  both models explain family replication\cite{Ng:1992st} and electric charge quantization\cite{deSousaPires:1998jc}}. Here we address a problem that only affect exclusively  case B that is a Landau-like pole that, as showed in previous studies, manifest already at TeV regime\cite{Ng:1992st,Dias:2004dc}. The most popular version of 3-3-1 models related to the case B is the minimal 3-3-1 model.

The landau pole arises in the minimal 3-3-1 model because the couplings $g_X$, associated to $U(1)_X$, and   $g_L$, associated to $SU(3)_L$, are related to  the Weinberg angle $\theta_W$ in the following way\cite{Ng:1992st,Dias:2004dc}
\begin{equation}
    \frac{g_X}{g_L}\sim \frac{\sin^2 \theta_W}{1-4\sin^2 \theta_W}.
    \label{coupling-relationI}
\end{equation}
This dangerous relation generates a Landau-polo,  $g_X(\Lambda) \rightarrow \infty$, when $\sin^2\theta_W(\Lambda) \rightarrow 1/4$. It was showed in Refs. \cite{Ng:1992st,Dias:2004dc} that this happens  already at TeV scale more precisely  in the range  $\Lambda \approx 2-6$ TeV depending on the particle content that contribute to the running of the $\sin^2 \theta_W$. Current bound on $Z^{\prime}$ demands that the $SU(3)_L \times U(1)_X$ symmetry breaks spontaneously around 4 TeV\cite{Cao:2016uur}. This  means that, in its original version, the minimal 3-3-1 model practically lost its power of making prediction.

It is important to stress that the minimal 3-3-1 model does not accommodate neutrino masses\cite{Pires:2014xsa}, neither has a candidate for dark matter and faces difficult in accommodating the recent result of $g-2$ of the muon\cite{deJesus:2020ngn} and the  B anomalies\cite{CarcamoHernandez:2022fvl}. In view of all this  the model need to be modified in order to explains such set of points. Thus, it is natural to look for extensions of the model that provides an answer to such points and evade or shift the Landau pole to a harmless energy scale. Following this line of thought, it was showed in \cite{Dias:2004wk} that the addition of octet of leptons may evade the Landau pole or shift it to a harmless energy scale.

Here we follow this line of thought and address this issue with the addition of  scalar leptoquarks\cite{Pati:1973uk} to the minimal 3-3-1 model. Leptoquarks are very well motivated proposal of new physics that may manifest at TeV scale. It also has an important role in flavor physics\cite{Dorsner:2016wpm}. Here,  we perform an exhaustive  investigation of  the impact of scalar leptoquarks on the landau pole of the model.  We show that leptoquarks are as efficient as  octet of leptons to evade or shift the non-pertubative regime of the minimal 3-3-1 model to a harmless energy scale.

\section{Revisiting the problem}
\subsection{The particle content of the model}

In the minimal 3-3-1 model leptons are arranged in  triplet representation of $SU(3)_L$,
\begin{equation}
f_{aL}= \begin{pmatrix}
\nu_{a_L}     \\
\ell_{a_L}       \\
\ell^{c}_{a_R} \\
\end{pmatrix} \sim (1,3,0),
\label{lp-rep}
\end{equation}
with $a=1,2,3$ representing the three generations of leptons.

In the Hadronic sector, anomaly cancellation requires that  one family transforms differently from the other two. This fact allows three possibility of arrangements for the quark families. Here we chose the third generation coming in triplet and the other two coming in  anti-triplet representation of $SU(3)_L$, 

\begin{eqnarray}
&&Q_{i_L} = \left (
\begin{array}{c}
d_{i} \\
-u_{i} \\
d^{\prime}_{i}
\end{array}
\right )_L\sim(3\,,\,3^*\,,\,-1/3)\,,u_{iR}\,\sim(3,1,2/3),\,\,\,\nonumber \\
&&\,\,d_{iR}\,\sim(3,1,-1/3)\,,\,\,\,\, d^{\prime}_{iR}\,\sim(3,1,-4/3),\nonumber \\
&&Q_{3L} = \left (
\begin{array}{c}
u_{3} \\
d_{3} \\
u^{\prime}_{3}
\end{array}
\right )_L\sim(3\,,\,3\,,\,2/3),u_{3R}\,\sim(3,1,2/3),\nonumber \\
&&\,\,d_{3R}\,\sim(3,1,-1/3)\,,\,u^{\prime}_{3R}\,\sim(3,1,5/3),
\label{quarks-rep} 
\end{eqnarray}
where  $i=1,2$. The primed quarks are  heavy quarks. 

The gauge sector is composed by nine gauge bosons where four of them are the standard ones $A\,\,\,,\,\,\, W^{\pm}\,\,\,,\,\,\, Z^0$ and the other five are the typical 3-3-1 gauge bosons  $U^{\pm \pm} \,\,\,,\,\,W^{\prime \pm}\,\,,\,\,Z^{\prime}$.  

The original scalar sector of the  model involves three triplets and one sextet of scalars, namely
\begin{eqnarray}
&&\eta = \left (
\begin{array}{c}
\eta^0 \\
\eta^-_1 \\
\eta^{+}_2
\end{array}
\right )\sim \,\,(1\,,\,3\,,\,0),\,\rho = \left (
\begin{array}{c}
\rho^+ \\
\rho^0 \\
\rho^{++}
\end{array}
\right )\sim\,(1\,,\,3\,,\,1)\,\,,\,
\chi = \left (
\begin{array}{c}
\chi^- \\
\chi^{--} \\
\chi^{ 0}
\end{array}
\right )\sim(1\,,\, 3\,,\,-1).\,\nonumber\\
&&\,\,\,\,\,\,\,\,\,\,\,\,\,\,\,\,\,\,\,\,\,\,\,\,\,\,\,\,\,\,\,\,\,\,\, S=\left(\begin{array}{ccc}
\, \Delta^{0} & \Delta^{-} & \Phi^{+} \\
\newline \\
\Delta^{-} & \, \Delta^{--} & \Phi^{0} \\
\newline \\
\Phi^{+} & \Phi^{0} & \, H_2^{++} \end{array}\right)\sim(1\,,\,6\,,\,0).
\label{scalar-cont} 
\end{eqnarray}
 When $\chi^0$ develops vacuum expectation value(VEV) different from zero, $v_\chi$, the $SU(3)_L \times U(1)_X$ symmetry breaks to the $ SU(2)_L \times U(1)_Y$ one. When the others neutral scalars develop VEV different from zero, the standard symmetry is spontaneously broken to the electromagnetic one. Such scalar content  generate masses for all massive particles of the model, except neutrinos.

After symmetry breaking $Z^{\prime}$ acquires the following mass expression\cite{Ng:1992st}
\begin{equation}
    M^2_{Z^{\prime}} \approx \frac{g^2\cos^2 \theta_W}{3(1-4\sin^2 \theta_W)} v^2_\chi.
    \label{primemass}
\end{equation}
The other terms are proportional to the electroweak scale and can be neglected. Current collider bound on $Z^{\prime}$  imposes $M_{Z^{\prime}}> 5$ TeV which implies $v_\chi > 4.3$ TeV\cite{Cao:2016uur}. As we show below, the model has a Landau pole that will manifest at some energy scale $\Lambda$. If $\Lambda <  4.3$ TeV  than the model is not predictive at all. 
\subsection{Landau-Pole}

In order to study the behaviour of the Weinberg angle with energy we, first, need to know how gauge couplings run with energy. In general the running of gauge couplings at one-loop is dictated by the relation
\begin{equation}
    \frac{1}{\alpha(\Lambda)_i}=\frac{1}{\alpha(\mu)_i}+\frac{1}{2\pi}b_i\log(\frac{\mu}{\Lambda}),
    \label{run}
\end{equation}
where $\alpha_i=\frac{g^2_i}{4 \pi}$. The renormalization coefficients for a general $SU(N)$ gauge group are given by \begin{equation}
    b_i=\frac{2}{3}\sum_{fermions}Tr(F)_i+\frac{1}{3}\sum_{scalars}Tr(S)_i-\frac{11}{3}C_{2}(G)_i.
    \label{coef}
\end{equation}
For $SU(N)$ we have $T_R(F,S)=1/2$ and $C_2(G)=N$.  For $U(1)$ we have $C_2(G)=0$. We also use $\sum Tr(F,S)=\sum y^2$ for $U(1)_y$ with $y=\frac{Y}{2}$, for the standard model case, and $y=X$ for the 3-3-1 case.

The running of $\sin^2 \theta_W(\Lambda)$ for $\Lambda < \mu_{331}$, where $\mu_{331}=\langle \chi ^0 \rangle=\frac{v_\chi}{\sqrt{2}}$,  is given by
\begin{equation}
    \sin^2 \theta_W(\Lambda)=\frac{1}{1+\frac{\alpha_2(\Lambda)}{\alpha_1(\Lambda)}}
    \label{angle1}
\end{equation}

\begin{figure}[t]
\centering
\includegraphics[width=0.9\columnwidth]{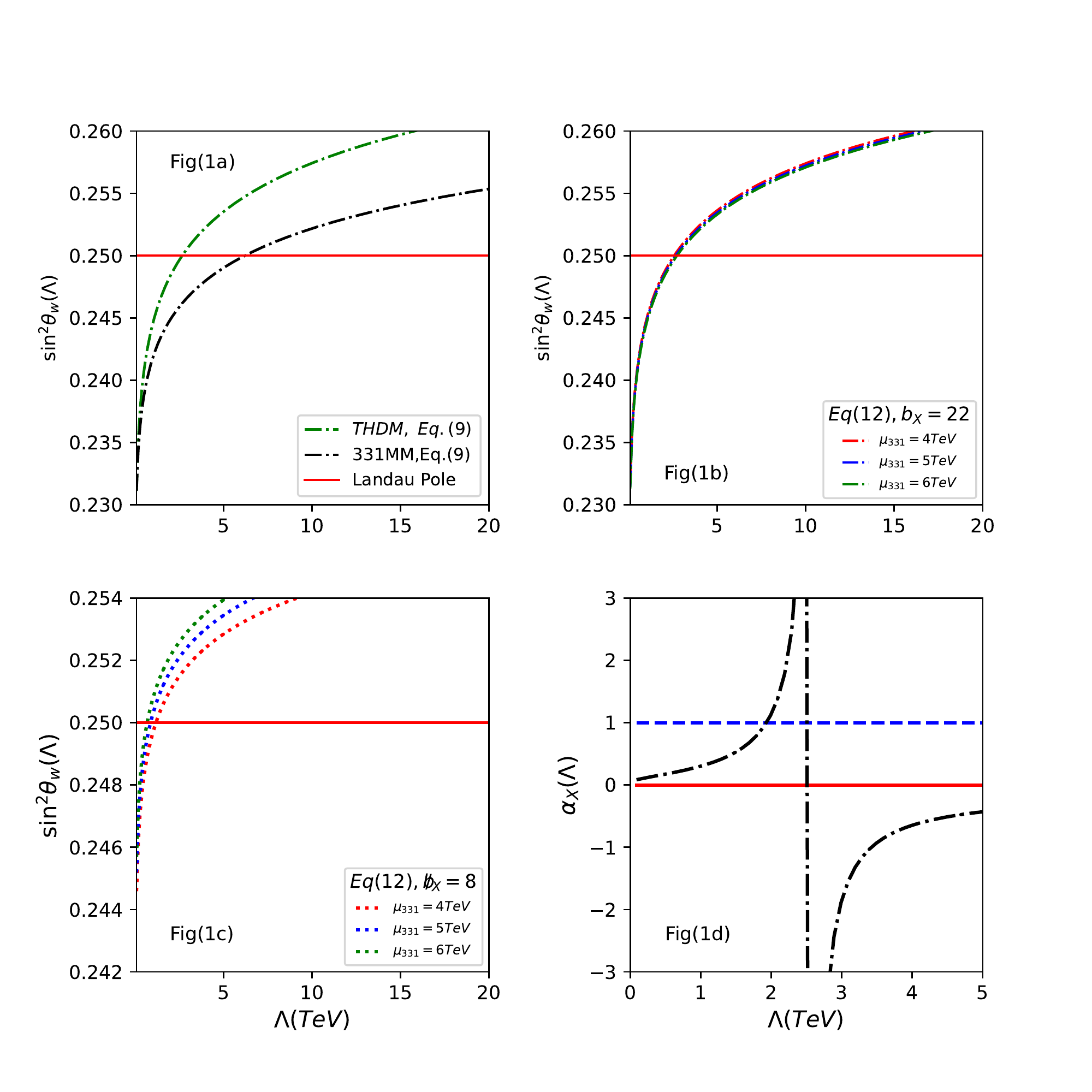}
\caption{ In this figure we show the running of the electroweak mixing angle , given by Eq.(\ref{angle1}), Fig.(1a), and  Eq.(\ref{angle2}) that leads to  Figs.(1b-1c). In the figure (1d) we show the  $\alpha_X(\Lambda)$ running to $\mu_{331}$ = 4TeV assuming  all particle content of the model. The contextualization of the curves behavior are described in the text.}
\label{fig1}
\end{figure}

For we have an idea about the energy scale the Landau pole arises, let us consider the simplest case in which the scalar sector of the  model is composed only with  three triplets of scalars. In this case, when of the $SU(3)_L \times U(1)_N$ symmetry breaking, the scalar sector decouples into the effective two Higgs doublet model (THDM) plus a set of scalars with mass belonging to the 3-3-1 scale. We, then, consider only the effective THDM which means the triplet $\chi$ and the singlets $\rho^{++}$ and $\eta_2^+$ and the exotic quarks  are not active degrees of freedom. In this case we get
\begin{equation}
    b_1=\frac{20}{9}N_F+\frac{1}{6}N_S=7\,\,\,\,\,\mbox{and} \,\,\,b_2=\frac{2}{3}N_F+\frac{1}{6}N_S-\frac{22}{3}=-3,
    \label{bpred}
\end{equation}
where $N_F$ is the number of fermion families, and $N_S=2$, is the number of scalar doublets.  Our results are displayed in FIG. 1.

In this simple case the running of $\sin^2 \theta_W(\Lambda)$ for $\Lambda < \mu_{331}$ is displayed in Fig.(1a), see dot-dashed line in green, where the landau pole corresponds to $\Lambda \sim 3.5TeV$.  We run the Eqs. (\ref{run}) and  (\ref{angle1}) considering the following values at the $M_Z$ energy scale $\sin^2 \theta _W(M_{Z}) = 0.2311$ , $\alpha_1(M_Z) = 1/128$ , $M_Z = 91.188GeV$ and $\alpha_1(M_Z) = \alpha_2(M_Z)\tan^2 \theta _W(M_{Z})$.

In the Fig. (1a), dot-dashed line in black,  we  considered  the scalar sextet S\footnote{ This scalar  decouples in one triplet with $Y=-2$, one iso-doublet with $Y=1$ and one doubly charged scalar with $Y=2$},  which implies an additional iso-doublet and a non-hermitian triplet, so that $N_S = 3$ and the addition  $N_T = 1$  that leads to 
\begin{equation}
    b_1=\frac{20}{9}N_F+\frac{1}{6}N_S + N_T = \frac{49}{6}   \,\,\,\,\,\mbox{and} \,\,\,b_2=\frac{2}{3}N_F+\frac{1}{6}N_S + \frac{2}{3}N_T -\frac{22}{3}= - \frac{13}{6}.
    \label{bpred}
\end{equation}
In this case the pole was pushed up a little bit,  to the value $\Lambda \sim 6TeV$ which means that the sextet of scalars is sufficient to recover the pertubative regime of the model concerning current bounds.


Let us now consider that all particle content of the model, the three triples, the sextet of scalars and the exotic quarks  are active degrees of freedom which means that  their masses  are below $\mu_{331}$.

To the case of energies above the scale  $\mu_{331}$,  the running of $\sin^2 \theta_W(\Lambda)$ for $\Lambda > \mu_{331}$ is 
\begin{equation}
    \sin^2 \theta_W(\Lambda)=\frac{1}{4(1+\frac{\alpha_L(\Lambda)}{4\alpha_X(\Lambda)})}
    \label{angle2}
\end{equation}
where now $\alpha_X(\Lambda)$  running equation we will be given by
\begin{equation}
    \frac{1}{\alpha_X(\Lambda)}=\left(1 - 4\sin^2 \theta _W(M_{Z})\right)\frac{1}{\alpha(M_Z)} +\frac{1}{2\pi}\left(b_1 - 3b_2\right)\log(\frac{M_Z}{\mu_{331}}) + \frac{1}{2\pi}b_{X}\log(\frac{\mu_{331}}{\Lambda} ),
    \label{running}
\end{equation}
and $\alpha_L(\Lambda =\mu_{331}) = \alpha_2(\mu_{331})$. In the equation above $b_X$ is the renormalization coefficient for $U(1)_X$. When  the degrees of freedom   above $\mu_{331}$ are taken into account we have $b_X = 20 + N_{\rho} + N_{\chi}$. Furthermore, when exotic quarks are omitted, we can introduce the notation ${b\!\!\!/}_X = 6 + N_{\rho} + N_{\chi}$, where now ${b\!\!\!/}_X = 8$. 

 Our results are showed in  Fig.(1b).  The dot-dashed line in red  corresponds to the case where $\mu_{331}=4TeV$, while the blue line to $\mu_{331}=5TeV$  and  finally the green line to $\mu_{331} = 6TeV$. In this figure we considered  all particle content of the minimal 3-3-1 model. This case 
leads to $b_X = 22$ and presents a Landau pole around (2-2.5) TeV which is below 4.3 TeV. This meas that in this case the model is no predictive at all. 

In Fig(1c) contributions of the exotic quarks  are omitted ( which gives ${b\!\!\!/}_X = 8$). We also assumed  the same choices for $\mu_{331}$ as in Fig.(1b). This corresponds to a more restrictive case for $\mu_{331} > 4TeV$, which implies the existence of a Landau pole to  $\Lambda < 2 TeV$. We see here that the Landau Pole is sensitive to the fact if the exotic quarks are active degrees of freedom or do not and to the  $\mu_{331}$ energy scale.

Finally, in Fig(1d) we present the behavior of the running of
$\alpha_X(\Lambda)$  given by Eq.(\ref{running}) to $\mu_{331}$ = 4TeV assuming  all particle content of the model. The figure let clear the position of the Landau pole at $\sim 2.5$ TeV and indicates the loss of the perturbative character of the model already in $\sim 2$ TeV which is very close to the Landau pole. In what follow we just discuss the Landau pole. 


We made here a short  review of the problem  concerning the Landau pole that arises in the minimal 3-3-1 model. The results we obtained is in agreement with the previous one. Conjugating this results with the current bound on the scale of the 3-3-1 symmetry breaking , given by $\mu_{331}=\langle \chi ^0 \rangle=\frac{v_\chi}{\sqrt{2}}$, which is around $4.3$ TeV, we conclude that the perturbative regime of the  minimal 3-3-1  model in its original form depends strongly if the exotic quarks are active degrees of freedom or do not. Even in this case the model is predictive up to the energy scale of 6 TeV, only, which is very close of $ 4.3$ TeV.

In what follow we make an exhaustive investigation of  the unique proposal existent in the literature that bring the model to the game by evading the Landau pole with a particular extension of the particle content of the model.

\subsection{ Evading the pole with octet of leptons}

It was proposed in \cite{Dias:2004wk} that  we could evade the pole in adding three octet of leptons to the minimal 3-3-1 model content. The octet   is composed by the following leptons
\begin{eqnarray}
\Xi=\left(\begin{array}{ccc}
\, \frac{1}{\sqrt{2}}t^0+\frac{1}{\sqrt{6}}\lambda^0 & t^+ & \delta^- \\
\newline \\
 t^-& \,-\frac{1}{\sqrt{2}}t^0+\frac{1}{\sqrt{6}}\lambda^0  & \delta^{--} \\
\newline \\
 \xi^{++}& \xi^{++} &  \, -\frac{2}{\sqrt{6}}\lambda^0  \end{array}\right)\sim(1\,,\,8\,,\,0).
\label{octeto} 
\end{eqnarray}
\begin{figure}[h]
\centering
\includegraphics[width=0.9\columnwidth]{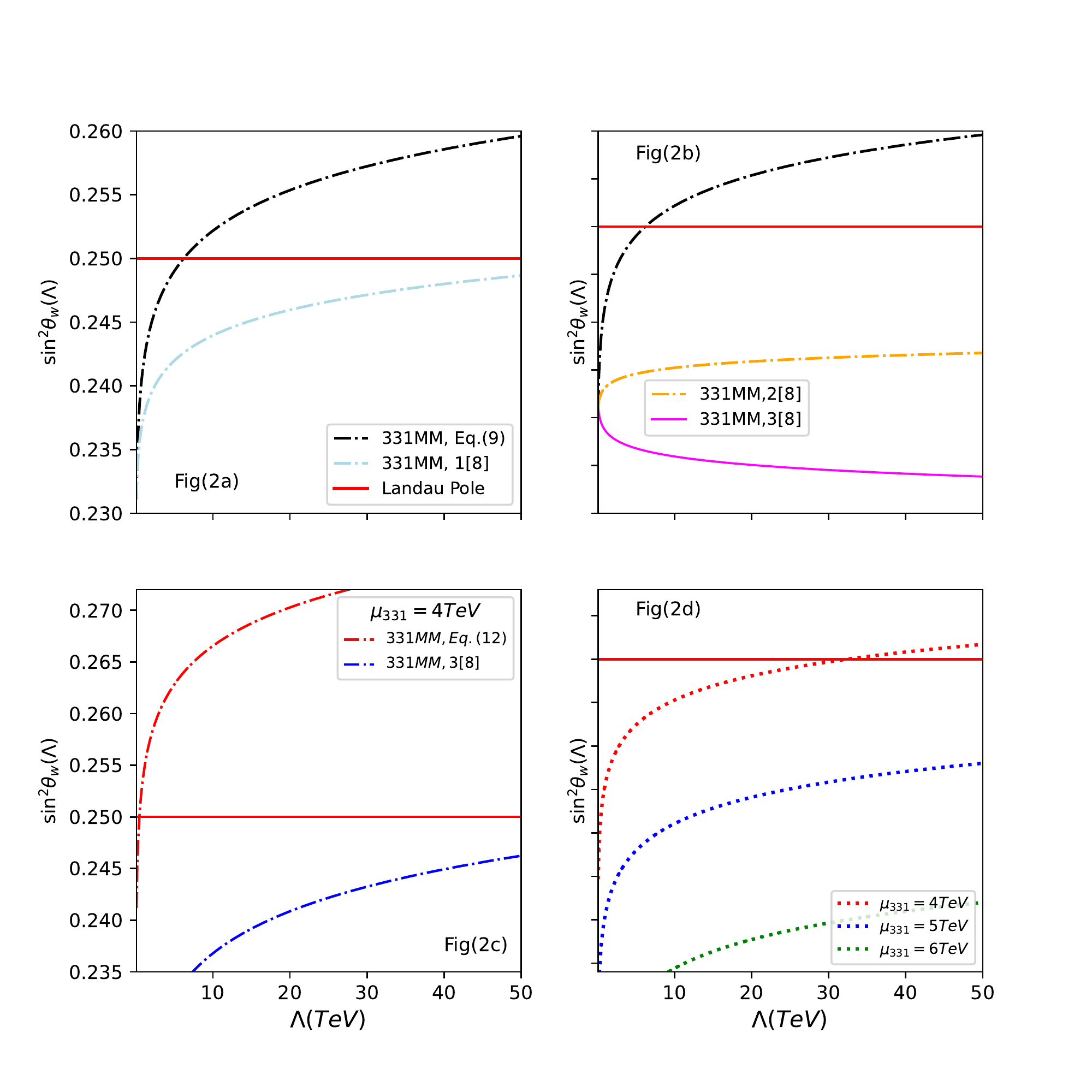}
\caption{ In this figure we show the running of the electroweak mixing angle , given by Eq.(\ref{angle1}), Figs.(2a-2b), and Eq.(\ref{angle2}) that leads to  Figs.(2c-2d). The contextualization of the curves behavior are described in the text.}
\label{fig1}
\end{figure}

Here we calculate the contribution of the octet of leptons to the running of $\sin^2 \theta_W(\Lambda)$ given by Eq. (\ref{angle1}),
that correspond to  $\Lambda < \mu_{331}$, for the cases when we  add one, two or three octets of leptons. In FIG. 2 we show our results. In the Fig. (2a) we  have the running  of the minimal 3-3-1 model (dot-dashed line in black) and, the dot-dashed light-blue curve, we 
have the running of the minimal 3-3-1 model+1-octet of leptons(1[8]). Observe that the addition of 1-octet of leptons is sufficient to shift the landau pole to energies above 50 TeV. In Fig(2b) we have the running  for the cases of
two and three octets of leptons.  

 When  the degrees of freedom above $\mu_{331}$ are considered, which means    all particle content of the model is took into account,  the running of $\sin^2 \theta_W(\Lambda)$ is given by Eq. (\ref{angle2}). The results for this case is displayed in  Fig(2c) where dot-dashed line in blue  corresponds  to  tree octets and  $\mu_{331}= 4TeV$.  As we can see in this case   the landau pole is shifted to $\Lambda > O(50) TeV$. 

For sake of completeness, in Fig.(2d) we present the running of the Weinberg angle for three values of $\mu_{331}$ when  the contributions of the exotic quarks  are omitted which means ${b\!\!\!/}_X = 8$, where only in this plot we  consider the addition of one octet. We see that for $\mu_{331} > 4$ TeV we have the absence of the Landau pole just for one octet of leptons. In general the higher  $\mu_{331}$ is, the harmless  the pole get. So we conclude that octet of leptons are efficient in circumventing the landau pole. Next we analyse other possibilities potentially interesting that evade the landau pole.  

\section{Evading the landau pole with leptoquarks }
In this section we consider the contribution of leptoquarks on the running of the Weinberg angle $\sin ^2\theta_W(\Lambda)$. Leptoquarks are very well motivated form of new physic that  is expect to manifest at TeV scale and then engenders  an interesting flavor physics scenario\cite{Dorsner:2016wpm}, give robust contributions to the $g-2$ of the muon and may generate neutrino masses at 1-loop\cite{Babu:2020hun,Parashar:2022wrd}. Moreover, leptoquarks may be probed at LHC\cite{Kramer:2004df}. 

Due to the the fact that the Yukawa interactions in 3-3-1 models discriminate family\cite{Ng:1992st,Oliveira:2022vjo}, i.e., one family of quarks must transform differently from the other two, then we are going to have a proliferation of leptoquark multiplets. To see this, observe that from the quark and lepton content of the minimal 3-3-1 model we can have scalar leptoquarks in the following representations,
\begin{eqnarray}
    && \bar L^C_{a_L} Q_{3_L} \sim (1\,,\,3\,,\,0)\times (3\,,\,3\,,\,2/3) \sim (3\,,\,3^* \oplus 6\,,\,2/3),\nonumber \\
    &&\bar L_{a_L} Q_{3_L} \sim (1\,,\,3^*\,,\,0)\times (3\,,\,3\,,\,2/3) \sim (3\,,\,1 \oplus 8\,,\,2/3),\nonumber \\
    &&\bar L^C_{a_L} Q_{i_L} \sim (1\,,\,3,0)\times (3\,,\,3^*\,,\,-1/3) \sim (3\,,\,1 \oplus 8\,,\,-1/3),\nonumber \\
    &&\bar L_{a_L} Q_{i_L} \sim (1\,,\,3^*,0)\times (3\,,\,3^*\,,\,-1/3) \sim (3\,,\,3 \oplus 6^*\,,\,-1/3),\nonumber \\
    &&\bar L_{a_L} d_{a_R} \sim (1\,,\,3^*,0)\times (3\,,\,1\,,\,-1/3) \sim (3\,,\,3^*_d\,,\,-1/3),\nonumber \\
    &&\bar L_{a_L} u_{a_R} \sim (1\,,\,3^*,0)\times (3\,,\,1\,,\,2/3) \sim (3\,,\,3^*_u\,,\,2/3)
    \end{eqnarray}
There are also the singlet leptoquarks that we do not consider in this work. The leptoquarks we are interested are these ones
\begin{eqnarray}
  &&  \phi^8_a \sim (3\,,\, 8\,,\, 2/3),\,\,\,\,\,\,\,\,\,\Phi^8_a \sim (3\,,\, 8\,,\, -1/3),\nonumber \\
  && \phi^6_a \sim (3\,,\,6\,,\,2/3),\,\,\,\,\,\,\,\,\,\, \Phi^6_a \sim (3\,,\,6^*\,,\,-1/3),\nonumber \\ &&  \phi^3_a \sim(3\,,\ 3^*\,,\,2/3),\,\,\,\,\,\,\,\,\,\,\Phi^3_a \sim (3\,,3\,,\,-1/3),
  \label{LQrep}
\end{eqnarray}
The indice $a$ refer to color. After symmetry breaking, these multiplet decompose as 
\begin{eqnarray}
    &&[8]_{X=2/3}=[3]_{Y=4/3}+[2]_{Y=-5/3}+[2^{\prime}]_{Y=13/6}+[1]_{Y=4/3}\nonumber \\
    && [8]_{X=-1/3}=[3]_{Y=-2/3}+[2]_{Y=7/3}+[2^{\prime}]_{Y=-11/3}+[1]_{Y=-2/3}\nonumber \\
    && [6]_{X=2/3}=[3]_{Y=-2/3}+[2]_{Y=7/3}+[1]_{Y=16/3}\nonumber \\
    && [6]_{X=-1/3}=[3]_{Y=-4/3}+[2]_{Y=5/3}+[1]_{Y=14/3}\nonumber \\
    && [3_d]_{X=2/3}=[2]_{Y=-2/3}+[1]_{Y=7/3}\nonumber \\
    &&[3_u]_{X=-1/3}=[2]_{Y=-5/3}+[1]_{Y=5/3},
    \label{decompo}
\end{eqnarray}
where $Y$ refers to the hypercharges of the leptoquarks .


Similarly to the previous  cases discussed above, here we calculated the contributions of these leptoquarks  to the running of $\sin^2 \theta_W(\Lambda)$ given by Eqs.(\ref{angle1}) as well by Eq.(\ref{angle2}) always having in mind economical scenarios. Our results are displayed in FIG. 3.
\begin{figure}[t]
\centering
\includegraphics[width=0.9\columnwidth]{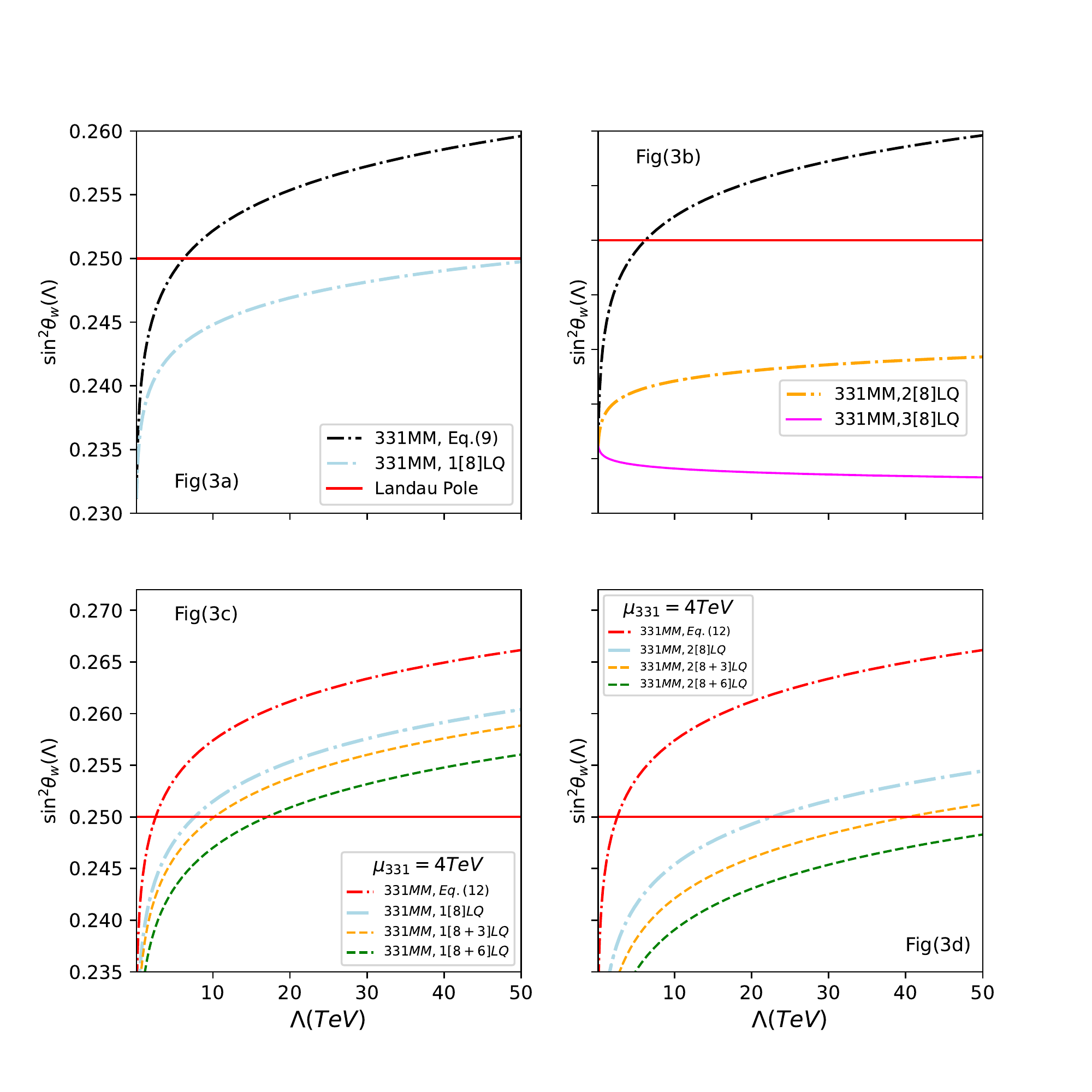}
\caption{ In this figure we show the running of the electroweak mixing angle , given by Eq.(\ref{angle1}), Figs.(3a-3b), and Eq.(\ref{angle2}) that leads to  Figs.(3c-3d). The contextualization of the curves behavior are described in the text.}
\label{fig1}
\end{figure}

 In the Fig. (3a), in the dot-dashed light-blue curve, we  assumed the addition of one octet of leptoquarks ($\phi^8_a $) to the minimal 3-3-1 model. In this case  the Landau pole is shifted to $\Lambda \sim 50 TeV$, while  the case for two and three octet of leptoquarks is depicted in the Fig(3b) where the dot-dashed orange line correspond to two octets of leptoquarks ($\phi^8_a, \Phi^8_a$) and the  magenta  to tree octets, and again the Landau pole is avoided at $\Lambda < \mu_{331}$. Observe that octet of leptoquarks  have the same effect than the octet of leptons discussed above and presented in Fig.2a and Fig.2b. 

However, when we  consider all degrees of freedom above $\mu_{331}$, i.e, when we  consider that all particle content of the model are active, in this case the running of $\sin^2 \theta_W(\Lambda)$ is given by Eq.(\ref{angle2}) and  the results are presented in  Fig(3c) with the dot-dashed line in red  corresponding again to  the minimal 3-3-1 model at $ \mu_{331}=4TeV$ with $b_X =22$. In this case  the addition of one octet of leptoquarks( dot-dashed light-blue curve) is sufficient to push the Landau pole  to values above 5 TeV which is a harmless scale according to current bounds. The other cases, as the addition of one  triplet or one sextet to the octet of leptoquarks  ($\phi^8_a, \phi^{(3,6)}_a$), are considered(dashed orange and green curves) and the Landau pole is shifted to $\Lambda \sim O(11-20) TeV$  which is a harmeless scale, too. In the  Fig(3d) we present the most general scenario involving leptoquarks  with hipercharges $(2/3, -1/3)[\phi,\Phi]$.

\section{Conclusions}
In this work we calculated the contributions of leptoquarks to the running of $\sin^2 \theta_W$ with the aim of obtaining their impact on the non-perturbative regime  of the model. The non-perturnative regime (a  Landau-like pole) of the model may manifest, depending on the particle content that contribute to the running of $\sin^2 \theta_W$,  already at few TeVs. Current bounds demand the $SU(3)_L \times U(1)_X$ symmetry  breaks  around 4 TeVs leaving the model, in its original form, phenomenologicaly  unpredictable.  Even in its original form, without the sextet of scalars, the model get non-perturbative below 4 TeV. The presence of the sextet, which is necessary to generate lepton masses, push this value to 6 TeV.

We have to resort to extension of the model in order to evade the non-perturbative regime and then recover its predictability. Thinking in this way, the addition of octet of leptons to the particle content of the minimal 3-3-1  model can do the job, as we can see in FIG. 2. In general, when we  consider that all the particle content of the model, including the three octet of leptons, contribute to the running of $\sin^2 \theta_W$ we have that the model predicts a Landau pole around $\Lambda \sim 100$ TeV.  The Landau pole is completely evaded only when we consider the contributions of  three octet of leptons in an effective 2HDM scenario.

In this work we analyzed the capacity of leptoquarks in evading the Landau pole. Our  results are displayed in FIG. 3. Firstly, the model allows leptoquarks in the representation of octets, sextets, triplets and singlets. As main result we showed that  leptoquarks are so efficient in evading the Landau pole as octet of leptons. For example, in adding leptoquarks to the scenario of 2HDM, one or two octet of leptoquarks are sufficient to shift the Landau pole to a harmless energy scale while the case of three octet evade completely the Landau pole. When we consider all particle content  active in the running of $\sin^2 \theta_W$, the most interesting scenario we found was the case of one octet plus one sextet of leptoquarks which pushed the Landau pole to an energy scale around 20 TeV. However, if we wish to push up the Landau pole  we may resort to the combinations presented in Fig.3d where the case of two octets and two sextet are very efficient pushing this energy scale to $\Lambda \sim 70$ TeV.

In concluding, we have that octet of leptons or multiplets of leptoquarks, both, are very efficient in evading the Landau pole that arises in in the minimal 3-3-1 model at  TeV scale. The advantage of leptoquarks are in their phenomenology once they are very attractive concerning flavor physics.

\section*{Acknowledgments}
C.A.S.P  was supported by the CNPq research grants No. 311936/2021-0 and A. Doff.  was supported by the CNPq research grants No. 310015/2020-0.
\bibliography{bibliography}
\end{document}